\documentclass[9pt]{entcs-local}

\usepackage{graphicx}
\usepackage{multirow}
\usepackage{longtable}
\usepackage{amsmath}
\usepackage{xifthen}
\usepackage{todonotes}
\usepackage{tikz}
\usepackage{entcsmacro-local}
\usepackage{subcaption}
\usepackage{mathtools}
\def\lastname{Arnaboldi, Czekster, Metere \& Morisset}

\newcommand{\observation}[1]{\todo[inline,size=\small,color=black!20,bordercolor=black!20]{#1}}
\newcommand{\obs}[2][]{{\ifthenelse{\isempty{#1}}{\observation{#2}}{\observation{#1: #2}}}}

\newcommand{\set}[1]{{\left\{ #1 \right\}}}

\begin{document}
\begin{frontmatter}
  \title{Modelling Load-Changing Attacks in Cyber-Physical Systems}
  \author{Luca Arnaboldi\thanksref{ALL}\thanksref{l.arnaboldi@newcastle.ac.uk}},
  \author{Ricardo M. Czekster\thanksref{ricardo.melo-czekster@newcastle.ac.uk}},
  \author{Charles Morisset\thanksref{charles.morisset@newcastle.ac.uk}}
  \address{School of Computing, Newcastle University\\
    Newcastle upon Tyne \\ United Kingdom}
  \author{Roberto Metere\thanksref{roberto.metere@newcastle.ac.uk}}
  \address{Newcastle University \& The Alan Turing Institute  \\
     Newcastle upon Tyne, London\\ United Kingdom}
    \thanks[ALL]{Order of authors is alphabetical.} 
    
    \thanks[l.arnaboldi@newcastle.ac.uk]{
    	Email: \href{mailto:l.arnaboldi@newcastle.ac.uk} 
    	{\texttt{\normalshape l.arnaboldi@newcastle.ac.uk}}}
    	
    \thanks[ricardo.melo-czekster@newcastle.ac.uk]{
    	Email: \href{mailto:ricardo.melo-czekster@newcastle.ac.uk} {\texttt{\normalshape ricardo.melo-czekster@newcastle.ac.uk}}}

    \thanks[roberto.metere@newcastle.ac.uk]{
    	Email: \href{mailto:roberto.metere@newcastle.ac.uk} 
    	{\texttt{\normalshape roberto.metere@newcastle.ac.uk}} }
    	
	\thanks[charles.morisset@newcastle.ac.uk]{Email:
	\href{mailto:charles.morisset@newcastle.ac.uk} {\texttt{\normalshape
		charles.morisset@newcastle.ac.uk}}}
		
\begin{abstract}

Cyber-Physical Systems (CPS) are present in many settings addressing a myriad of purposes.
Examples are Internet-of-Things (IoT) or sensing software embedded in appliances or even specialised meters that measure and respond to electricity demands in smart grids. 
Due to their pervasive nature, they are usually chosen as recipients for larger scope cyber-security attacks. 
Those promote system-wide disruptions and are directed towards one key aspect such as confidentiality, integrity, availability or a combination of those characteristics. 
Our paper focuses on a particular and distressing attack where coordinated malware infected IoT units are maliciously employed to synchronously turn on or off high-wattage appliances, affecting the grid's primary control management. 
Our model could be extended to larger (smart) grids, Active Buildings as well as similar infrastructures. 
Our approach models Coordinated Load-Changing Attacks (CLCA) also referred as GridLock or BlackIoT, against a theoretical power grid, containing various types of power plants. 
It employs Continuous-Time Markov Chains where elements such as Power Plants and Botnets are modelled under normal or attack situations to evaluate the effect of CLCA in power reliant infrastructures. 
We showcase our modelling approach in the scenario of a power supplier (e.g. power plant) being targeted by a botnet. 
We demonstrate how our modelling approach can quantify the impact of a botnet attack and be abstracted for any CPS system involving power load management in a smart grid. 
Our results show that by prioritising the type of power-plants, the impact of the attack may change: in particular, we find the most impacting attack times and show how different strategies impact their success. 
We also find the best power generator to use depending on the current demand and strength of attack.
\end{abstract}

\end{frontmatter}

\section{Introduction}\label{sec:intro}
Cyber security is a major concern when evaluating critical resource infrastructures. 
Malicious attacks and unintended damages significantly increase each year, and it is therefore important to know what effects they will have.
Due to strict supply-demand requirements in power grids, to maintain equilibrium is of paramount importance.
In practice, the entities that regulate and those that actually maintain equilibrium depend on the politics of the country, and we do not refer to them specifically, rather we refer to a {Cyber-Physical System} (CPS) component called {Load Controller}.
Whenever it is required to increase or decrease energy levels, it may trigger costly responses, e.g. turning on new energy sources or disconnecting areas from the grid.
Historically, power generation and demand have been very separated; however, this scenario is changing, and consequently their related security risks and possible attacks are changing too, so the current mitigation strategies may not be applicable any more.
The current role of a load controller is balancing energy supply accordingly to demand while maintaining the frequency of the current at around 50 MHz in Europe (or 60 MHz for other countries such as USA, Brazil and Japan).

Organised attacks aimed at power infrastructures are called {Coordinated Load-Changing Attacks} (CLCA), where synchronous connections or disconnections of high-wattage units such as water heaters or air conditioning units are used to cause disruptions in energy provision.
CLCA are here considered as \emph{black boxes}, i.e. they do not require extensive knowledge as to the particularities of grid operations in order to be employed.
If sudden spikes or drops such as synchronised turning on or off of several devices takes place, they cause the equipment to short and break, causing damages and reducing the availability of power supply.
The load controller can easily cope with common occurrences in terms of imbalances, adjusting energy flows accordingly; however, sudden usage spikes may unadvisedly cause the grid to collapse.
Malicious users could profit from those situations as they could infect a considerable number of high-wattage devices to coordinate actions that impairs energy distribution~\cite{dabrowski2017grid,dabrowski2018botnet}.

In the context of this paper, we are interested in two types of systems: on the one hand, we consider CPS~\cite{lee2008cyber,rajkumar2010cyber}, i.e. systems with limited resources (low power, energy, processing or other capacity related issue) employed in a variety of equipment ranging from sensors to smart grid components.
On the other hand, we are taking into consideration large scale infrastructures such as the {Smart Grid}, \emph{Active Buildings}, different types of power plant (that can be solar or nuclear, for example) and other power reliant schemes.
We are specifically focusing on modelling their interaction, i.e. how power supply mechanisms react due to CPS usage, as well as the influence one has on the another.
Through this model we can observe the daily usage of energy in the smart grid, and calculate its ability to cope with duress.
Our modelled attacker (or adversary) attempts to exploit the CPS mechanisms to decrease availability and cause damages; our model captures the probability of success and the tolerance of a theoretical smart grid made of various different types of power plants.

In this application context, it is worth defining a CPS and its roles.
It consists of components with two parts, i.e. a computing part integrated with a physical counterpart, both connecting and communicating with other CPS to achieve common objectives.
For example, a CPS could be an embedded system attached to a heartbeat sensor component to collect meaningful health data from patients inside a care unit, or could be smart meters exchanging data on power system energy transmission and distribution.
These infrastructures usually encompass a sizeable number of entities in hyper-connected environments, deployed to help users improve productivity, bottleneck assessments and much more.
The presence of general purpose CPS in residential, commercial and industrial settings is ubiquitous as many vendors offer solutions that vary from smart home sensing devices to closed circuit televisions.

The simplistic nature of CPS implementations makes them prone to software contamination: for instance, installing malware in high-wattage devices, e.g. air conditioning units or water heaters.
Due to this, they are natural recipients for coordinated attacks aimed to disrupt grid infrastructures as a synchronised event may cause over demand or influence voltage and frequency to inadmissible levels.
This is particularly unsettling for many reasons as it may impact overall energy costs for customers or, in extreme cases, lead to preventable casualties in healthcare settings.
The aim of this work is to evaluate different power plant configurations in terms of operational characteristics such as cooling down and maintenance time as well as type (e.g. nuclear power or solar panels), aiding managers to make better decisions when designing power systems, and potentially mitigating the impact of these attacks by choosing better load controller configurations.
Our model will also show the trade-offs of mixing different power plant types to withstand CLCA attacks, reducing the hazardous effects they have on the infrastructure.

The paper is organised as follows: Section~\ref{sec:security} will address cyber security concerns and related work.
On Section~\ref{sec:model} we will discuss our modelling, with our problem formalisation and a simple model.
We explain our cyber security model applied to CPS on Section~\ref{subsec:cps-model}.
Section~\ref{sec:results} will describe our results and findings and in Section~\ref{sec:final} we will discuss final considerations, model extensions and future work.

\section{Security concerns and related work}\label{sec:security}

A {CPS} often entails limited resource machinery, equipped with wireless or wired communication capabilities.	
They serve distinct purposes and could be deployed in different settings to address a multitude of objectives.
In this work we are interested only in \emph{availability} of CPS and the consequences it imposes to overall users (or customers) when used as recipient for active attacks.
In performance evaluation research, availability is a crucial measure used to infer quality levels of systems as well as capacity planning.
Attacks directed at availability aim to exhaust resource capabilities until depletion causing delays while processing messages.
CPS often operate unsupervised in a state known as {Machine to Machine} interaction (M2M). 
Scenarios involving direct {M2M}, without human supervision, can have several security implications; it becomes a lot easier for an attacker to gain undetected access to devices, attacks on these systems take longer to be detected and the possibility of a device misbehaving (due to attacker control or other) becomes much more common.

The vast popularity of these kinds of systems has triggered unprecedented (and inexpensive) access to {Botnets}. 
Simply put, a Botnet is a network of infected devices that can be remotely and synchronously controlled by an attacker.
With the raise of popularity of the IoT, it has become almost trivial to gain access to large amount of devices.
This is an outcome of lack of security measures applied to IoT devices, as well as the ease of access to tools such {BlackEnergy}\footnote{T. Birdsong and G. Davis. Updated blackenergy trojan grows more powerful, McAfee Labs, Mar 2018,  Available Online at: \href{https://securingtomorrow.mcafee.com/other-blogs/mcafee-labs/updated-blackenergy-trojan-grows-more-powerful/} {\texttt{https://securingtomorrow.mcafee.com/other-blogs/mcafee-labs/updated-blackenergy-trojan-grows-more-powerful/}}}.
This tool in particular was able to create a Botnet that took down large amounts of the Ukraine power grid in 2015\footnote{K. Zetter. \emph{Inside the cunning, unprecedented hack of Ukraine's power grid}. Wired, Jun 2017, Available Online At: \href{https://www.wired.com/2016/03/inside-cunning-unprecedented-hack-ukraines-power-grid/} {\texttt{https://www.wired.com/2016/03/inside-cunning-unprecedented-hack-ukraines-power-grid/}}}.
{Denial-of-service} (DoS) is a common choice of attack after gaining access to a botnet.
These attacks are directed towards availability and aim to interrupt service to critical systems.
Traditionally, DoS attempts are network-based and are used to saturate communication channels with spurious traffic to overwhelm active nodes until they stop servicing or begin acting abnormally.
These kinds of attacks flood targets with fake requests that may cause extreme delays, reducing service levels on recipients.
One key aspect of such service disruptions is to differentiate active and aggressive attacks from actual requests (even for an elevated number of messages), responding equitably to incoming messages.

DoS can be caused by various techniques, such as {Ping Flood}, {Smurf Attack}, {SYN Attack}, {Buffer Overflow}, and other attack techniques.
However, more recently, attackers began exploiting other system specific aspects, the kinds of attacks do not merely attempt to flood a system but rather exploit a specific vulnerability to cause damage which is harder to trace.
An example of exploiting a specific vulnerability was the attack against the Ukrainian power load controller that caused a nationwide blackout. 
However, it was not an isolated incident: as recently as this year, a similar cyber attack was suspected to have been launched against the Russian power grid\footnote{BBC News. \emph{US and Russia clash over power grid 'hack attacks'}. BBC News, Jun 2019,  Available Online At: \href{https://www.bbc.co.uk/news/technology-48675203} {\texttt{https://www.bbc.co.uk/news/technology-48675203}}}.
The common goal of these attacks is to create an imbalance in the frequency appreciated by power generators' security mechanisms due to a sudden surge.
Those sort of attacks have been called {load changing Distributed DoS} or {Grid-Shock}~\cite{dabrowski2017grid}, and they can cause damage to machinery, power cut-outs and (consequently) huge monetary loss (another name for this sort of attack is \emph{Flash Attacks}~\cite{kumar2019detecting}).

In CPS, in order to replace the need for user intervention and regulate power more reasonably, load balancing algorithms are used instead.
Load-balancing algorithms distribute resources based on expected energy usage to optimise the consumption, focusing on delivering energy more efficiently.
However, if massive unexpected spikes are triggered as a consequence of a botnet, it can cause waste of resources and even a scarcity of power for essential functions and real users.

Load-Changing attacks are not only pertinent to power-grids, it is becoming more common for smart infrastructures, such as smart buildings and smart cities.
The core idea is to have self regulating power supplies that adjust energy levels due to expected consumption and loads.
What this means is that a targeted attack could be aimed at a specific company or location to turn off power and overpower the load management system.
What is made evident by these examples~\cite{dabrowski2017grid,soltan2018blackiot}, is that this threat is no longer a far-fetched scenario, but rather, a very real likelihood. 

As a response to such challenges in power related research we have created a stochastic model that can be used by smart infrastructure managers to evaluate the threat on this system and devise countermeasures to tackle disruptions.
Our proposed modelling approach observes the impact on security of different control mechanisms of a CPS grid under a botnet attack.
In particular, we show the likelihood of the success under different setups: a detailed analysis is presented in Section~\ref{sec:results}.

\subsection{Related work}\label{subsec:related-work}

DoS attacks have long been one of the most common and dangerous threats in many computer networks.
Their detection~\cite{carl2006denial} is therefore the first step required to perform an effective response.
These attacks become even more dangerous as the IoT spreads across a vast amount of spectra and parts of life.
The literature on security concerns highlights similar scenarios of DoS attacks against IoT systems and CPS~\cite{arnaboldi2018generating,liang2016denial,roman2013features,cardenas2019iot}
With the ever changing Internet landscape, attackers have started to focus on specific vulnerabilities of systems to optimise their attacks. 
Liang et al.~\cite{liang2016denial}, showed a simple Distributed DoS attack on an IoT scenario, however, they have demonstrated that the result of an attack, if propagated to a CPS, could be massively impactful. 
In their work, Roman et al., 2013~\cite{roman2013features} mention key features of how the way these type of systems are setup can cause security concerns.
The unique characteristics present in smart infrastructures may render it vulnerable to new avenues of attack such as battery drain and new types of DoS.
These new challenges raise concerns for security professionals such as what security vulnerabilities their specific system could be subject to, and what impact it might have.

Load changing attacks were mentioned in Dabrowski et al., 2017~\cite{dabrowski2017grid}. 
In their work the authors have discussed a simulation concerning the impact that load changing has on power management. 
Their attack is based on the fact that when operating a power grid, providers have to continuously maintain a balance between supply (i.e., production in power plants) and demand (i.e., power consumption) to maintain the power grid’s nominal frequency of 50/60 Hz. 
Their Matlab simulations show that this balance can easily be broken through a botnet attack.
Through their power analysis, they also estimate the number of devices needed to disrupt a country's power grid.
This work is one of the first to show that a potential attack can be staged against a power plant without the need for manipulating the controls itself, but just by external device activity.
Our proposed approach takes inspiration from this external influences, but scales it to the representation of any power grid.
Also, rather than using simulation, our model uses model checking features present in Markovian solvers (PRISM tool) to evaluate the impact these attackers have on the modelled system.

Another work which studied coordinated attacks by botnets and disruptions to the power grid  was done by Soltan et al., 2018~\cite{soltan2018blackiot}.
Their work introduces the concept of {Manipulation on Demand via IoT} (MADIoT), through simulations they show how external influences of high power devices can cause disruptions and power outages.
They have demonstrated the interdependence between supply of power and the demand, and how this can be exploited to cause disruptions.
These types of attack are the core focus of our work, however, Soltan and his colleagues focused on simulations of attacks.
In contrast, it is our wish to quantify what these attacks mean from the perspective of a potential supplier. 
Through model checking we could investigate how the attacks affect the power system, and we can adapt it to better adjust, prepare, or respond to these attacks.

{Continuous Time Markov Chains} (CTMC) models have been successfully used to simulate attacks on a variety of systems~\cite{arnaboldi2017quantitative}.
Baumann et al., 2012~\cite{baumann2012markovian}, make use of CTMCs to model Flooding DoS on a theoretical network.
Through their models they were able to show the impact of the attacks on the systems throughput and evaluate its effects.
They could also perform security checking for different DoS rates and scenarios.
In Arnaboldi \& Morisset, 2017~\cite{arnaboldi2017quantitative}, this notion is taken a step forward to model a CTMC capturing DoS attacks on IoT Systems.
The model presented were able to quantify several impacts of the DoS attacks including how they influenced other components in the system as well as suggesting optimal system setups.
Both these works demonstrate the flexibility of CTMCs for modelling attacks on systems of devices.
Our proposed modelling methodology uses a similar approach to modelling systems of devices and observes the impacts of the attacks, but rather than focusing on general flooding of messages looks at the specific problem of load balancing and power management.
Previous work in this area by Norman et al. 2005~\cite{norman2004power} used PRISM to observe runtime strategies in order to achieve a trade-off between the performance and power consumption of a system. Our approach extends this to look at the influence of an attacker on these balancing strategies

Specific to the context of coordinated cyber-attacks on smart grids, Moya \& Wang, 2018~\cite{moya2018develop} have developed correlation indices suitable for identification of these disruptions.
Sun et al., 2016~\cite{sun2016coord} have proposed a \emph{Coordinated Cyber Attack Detection System} (CCADS), strongly inspired in IDS concepts as well as its benefits to cyber security efforts to mitigate the effects of such disturbances in smart grids.

\subsection{Threat Model}

We are working on the assumption that our designed adversary has gained illicit access to a large number of IoT devices and formed a botnet.
Using this botnet, our envisioned attacker targets a smart grid infrastructure through excess energy usages and causes a spike which damages the load controller.
When a power plant experiences very high load, it will employ one out of three mitigation controls, primary, secondary or tertiary~\cite{dabrowski2017grid,soltan2018blackiot}. 
\begin{enumerate}
  \item \emph{Primary Control} distributes the load to other power-plants in the vicinity.
  \item \emph{Secondary Control} makes an assessment to return to normal operation if criteria is met.
  \item \emph{Tertiary Control} frees up resources from previous Primary and Secondary controls.
\end{enumerate}
The most damaging target is Primary Control, as switching on and turning off further plants is expensive and time intensive.
The objective of the attack is to continuously trigger the Primary control and consequently cause the most damage, including potential damage to turbines and machinery caused by the strain.

Further to this consideration, power suppliers are constantly balancing the frequency of the supply. 
Sudden spikes may affect the frequency significantly enough to activate security mechanisms of power plants for which they detach themselves from the grid.
This may lead to blackouts and disruptions.
A smart intruder who has gained access to a botnet can choose when to turn them all on synchronously.
If she controls enough devices, she may induce a spike and make suppliers detach.
In order to inflict the highest damage, the attacker needs to make sure that she will cause a spike; however depending on current usage, this may or may not happen.
If the attacker controls a fraction of the devices and turns them all on, but these devices are already operational due to expected daily usage, the spike may not trigger.
Perhaps counter-intuitively, depending on the daily usage, it might be a lot more damaging to trigger the spike at a lower usage time such as mid afternoon to cause the most disruption.

The effectiveness of a spike will also depend on the type of supplier providing the energy.
Whilst if a nuclear supplier is spiked it might take a very long time to recover, gas power plant have a much higher adjustment rate and are therefore more resilient to these attacks.
In our model we mimic the response behaviours of hydro, gas and nuclear power.
We model the demand borrowing real values\footnote{The demand across the UK of the 27 September 2019.} in the UK, which we scale down to limit the number of power suppliers in the model.
The attacker is modelled at every hour of the day (with the mean power usage at that time), and we calculate the success and impact rate of the attacks.

\section{Model}\label{sec:model}
{Markov Chains} (MC) are a powerful modelling formalism to describe behavioural properties of systems with simple primitives~\cite{norris1998markov} proposed in the early $20^{th}$ century by the mathematician Andrey Markov.
However, they do not appear in the context of time shared systems if not by mid $1960$ for scalability purposes~\cite{scherr1967analysis,stewart2007perf}.
The idea behind MC is to abstract a system using only states, transitions and rates or probabilities to model behaviour.
Solutions of Markovian based systems usually employ direct or iterative solvers, yielding state permanence probabilities.
It is possible to model situations where randomness representation is important; this has been successfully applied throughout the years in different domains such as economy, music composition, {information retrieval} and ecology.

Despite their strength, a known problem affecting MC based approaches concerns how easy one could end with an intractable model having millions of states.
This problem is known as {state space explosion}, where even small systems (for example, systems having a couple of limited capacity queues) may show an unmanageable number of states.
One way to mitigate state explosion is to resort to \emph{structured} alternatives that still work with an underlying MC, profiting from its modelling strength and operating with larger models.
Examples of formalism employing reduction strategies are {Queueing Networks} (QN)~\cite{stewart2009prob}, {Stochastic Petri Nets} (SPN)~\cite{molloy1982perf} as well as more modular ones such as {Performance Evaluation Process Algebra} (PEPA)~\cite{hillston2005compositional}, {Superposed Generalized Stochastic Petri Nets} (SGSPN)~\cite{donatelli1994super}, and {Stochastic Automata Networks} (SAN)~\cite{plateau1991stochastic}, to name a few examples.
It is worth mentioning that some approaches share many constructs among each other such as the concept of \emph{modules}, e.g., \emph{processes} in PEPA or \emph{automata} in SAN (in a very abstract perspective).

The present work aims to model a CPS based environment under attack employing {reactive modules}~\cite{alur1999reactive}, a modelling approach used in the {PRISM} tool~\cite{kwiatkowska2011prism}.
The software is a {Probabilistic Model Checker} used in verification, where modellers are able to choose from different types of probabilistic models such as {Continuous Time Markov Chains} (CTMC), {Discrete Time Markov Chains} (DTMC), {Probabilistic Automata} (PA), {Probabilistic Timed Automata} (PTA), and {Markov Decision Processes} (MDP) with extensions to models with costs and rewards.
A {graphical user interface} guides the modeller when building and analysing models, with interesting features such as {path analysis}, system properties, computation of steady and transient state probabilities, and a simulator used for model debugging.
It also allows for the definition of symbolic variables combined with experiments that operate under intervals and steps, easing analysis.
Internally, it employs solution methods such as Power, Jacobi, Gauss-Seidel methods (among others) for CTMCs and DTMCs.
For more information we refer to PRISM's website\footnote{PRISM: Probabilistic Symbolic Model Checker, online at \href{ https://www.prismmodelchecker.org} {\texttt{ https://www.prismmodelchecker.org}}}.
For the set of modelling possibilities stated earlier, PRISM allows for straightforward model decomposition into so called \emph{modules}, where users employ primitives such as global variables, state transitions with associated rates, formulas assigned to states (e.g. functions that observe local states in other modules or other desired behaviour), among others.

\subsection{Continuous Time Markov Chains}
\label{sec:ctmc}
A CTMC is a stochastic process having the {Markov property}~\cite{norris1998markov}, also known as {memoryless property}, that is usually defined as:
\[
P[X(t_k) = s_k ~|~ X(t_{k-1}) = s_{k-1},~\ldots~, X(t_{0})=s_{0} ] = P[X(t_k)=s_k ~|~ X(t_{k-1}) = s_{k-1}],
\]
where \{ $X(t) ~|~ t \in \mathbb{R}_{\geqslant0}$ \} are random variables, $X(t)$ are observations at time $t$ (e.g. $X(t)$ corresponds to the system at time $t$).
The memoryless property addresses the notion that in a given state, the decision of visiting a next state must not take into account the set of prior visited states, e.g, the past up to this point is irrelevant.
A CTMC shows similarities with a {Labelled Transition System} (LTS), but it is different as its transitions are decorated with exponentially distributed (memoryless) delays or rates instead of labels.
Another difference is that in LTS (also known as {Kripke structure} or {Finite Automata}) the labels on transitions allow the modelling of non-determinism, being a powerful formalism used in formal verification due to its high scalability and applicability~\cite{hartmanns2015quant}.

A CTMC assumes enumerable states where time evolves continuously according to the chosen values for the set of rates~\cite{kwiatkowska2011prism}.
Formally, a CTMC consists of a tuple $(S, s_{0}, R, L)$ where
\begin{itemize}
 \item $S$ is the state space of the problem, i.e., the finite set of states modelled by the user,
 \item $s_{0} \in S$ is the initial state,
 \item $R: S \times S \rightarrow \mathbb{R}_{\geqslant0}$ configures the transition rate matrix of the model, and
 \item $L: S \rightarrow 2^P$ are label mapping states to atomic propositions in $P$.
\end{itemize}
If multiple states have ongoing transitions with different rates, i.e. $R(s,s') > 0$, then a \emph{race condition} occurs, where the transition with the least value (according to the exponential distribution of the rate) is triggered first.
The time spent in a state is drawn from the exponential distribution given by
$E(s) = \sum_{s'} R(s,s')$, where $E(s)$ is the exit rate of the state $s$ and the probability of leaving a state $s$ within $[0,t]$ equals to $1-e^{-E(s)t}$.

A CTMC encloses a discrete time counterpart termed the {Embedded DTMC}, computed using the transition rate matrix elements and $E(s)$, i.e. dividing each cell element by $E(s)$.
The so called \emph{infinitesimal generator} $\mathbf{Q}$ of a CTMC is a matrix $S \times S$ whose non-diagonal entries are the same as in $R$, and the diagonal entries are defined in such a way that the sum of each row of $\mathbf{Q}$ equals $0$.

\begin{equation*}
  \mathbf{Q}(s,s') =
    \begin{cases}
      R(s,s') & s\neq s'\\
      \displaystyle-\sum_{s \neq s'} R(s,s') & \text{otherwise}
    \end{cases}.
\end{equation*}
This concerns the representation of CTMCs and its operational meaning.
In terms of solution, it corresponds to multiplying an initial probability vector (it could start with equiprobable values) $\pi$ by $\mathbf{Q}$ until the balance is reached, i.e., $\pi\textbf{Q}=0$.
If the system has reached convergence, it presents the steady state of the model, otherwise, it diverged, and other techniques such as transient analysis should be employed instead.
The steady state yields the \emph{permanence probabilities} for each state, representing the system's behaviour after a long run, i.e., its stationary behaviour.

As stated earlier, PRISM is inspired by the formalism of \emph{reactive modules} as the decomposition approach to modelling.
It consists on devising smaller CTMCs for each module and then define transition functions using formulas, as well as synchronisation among modules.
For solution, it combines the local state spaces for all modules (e.g. Cartesian product).
PRISM is equipped with several numerical methods (Power, Jacobi, Gauss-Seidel, and many others), simulation or, depending on the tractability of the model (due to state space explosion), through model verification of properties or probabilistic model checking.

\subsection{Problem formalisation}\label{subsec:formalisation}

We demonstrate our main idea through the definition of a small order CTMC that will guide our process when we extend it to the set of power plants, load controller and botnets.
We are working here with a very reduced state space, explaining our initial state, some possible transitions and what we actually mean by an attack.

In a simplistic view of our model, we have a single power generator (PG) that can be in three states $S_{PG}$: (i) available ($a$), ready to supply but not yet generating energy to the grid, (ii) generating ($g$), currently providing energy to the grid, or (iii) restart ($r$), detached from the grid, not generating, nor supplying, and in the need of a restart.
The states in $S_D$ model the demand $D$ loading the grid at the average expected level $m$ (for {\em medium}) plus or minus small deltas, modelled by two additional states $l$ (for low) and $h$ (for high).
The attacker $B$ is a botnet controlling a large amount of infected devices.
Its states in $S_B$ are modelled as simply $0$, when the infected controlled devices are all off, or $1$ when those devices are switched on.
In summary we have the following states
\begin{equation*}
  \begin{split}
    & S_{PG} = \set{a, g, d} \\
    & S_{D} = \set{l, m, h} \\
    & S_{B} = \set{0, 1} \\
  \end{split}
\end{equation*}

Following the notation introduced in Sec.~\ref{sec:ctmc}, the CTMC relating to this simplistic model is defined as a tuple $(S, S_0, R, \mathcal{L})$ where $S = S_{PG} \times S_D \times S_B$, and $s_0 = \left( a, m, 0 \right)$.
The rate matrix $R$ is a square matrix of dimension $|S|$ whose entries are zeros apart from those corresponding to the transitions we modelled.
So for example, given two states $s_i$ and $s_j$, then the transition rate $r_{ij}$ is the mean per time unit that we expect the transition from $s_i$ to $s_j$ to happen.
The labels in $L$ associate valid proposition to states: we associate them to desirable properties.
In particular, we label all states where the current demand in the system is above the current supply as {\em ``overDemand''}, for example $\left( d, m, 1 \right)$, where if the PG is detached it cannot match the demand $m$.
\begin{figure}
\centering
\begin{subfigure}{.3\textwidth}
  \centering
  \includegraphics[width=.9\linewidth]{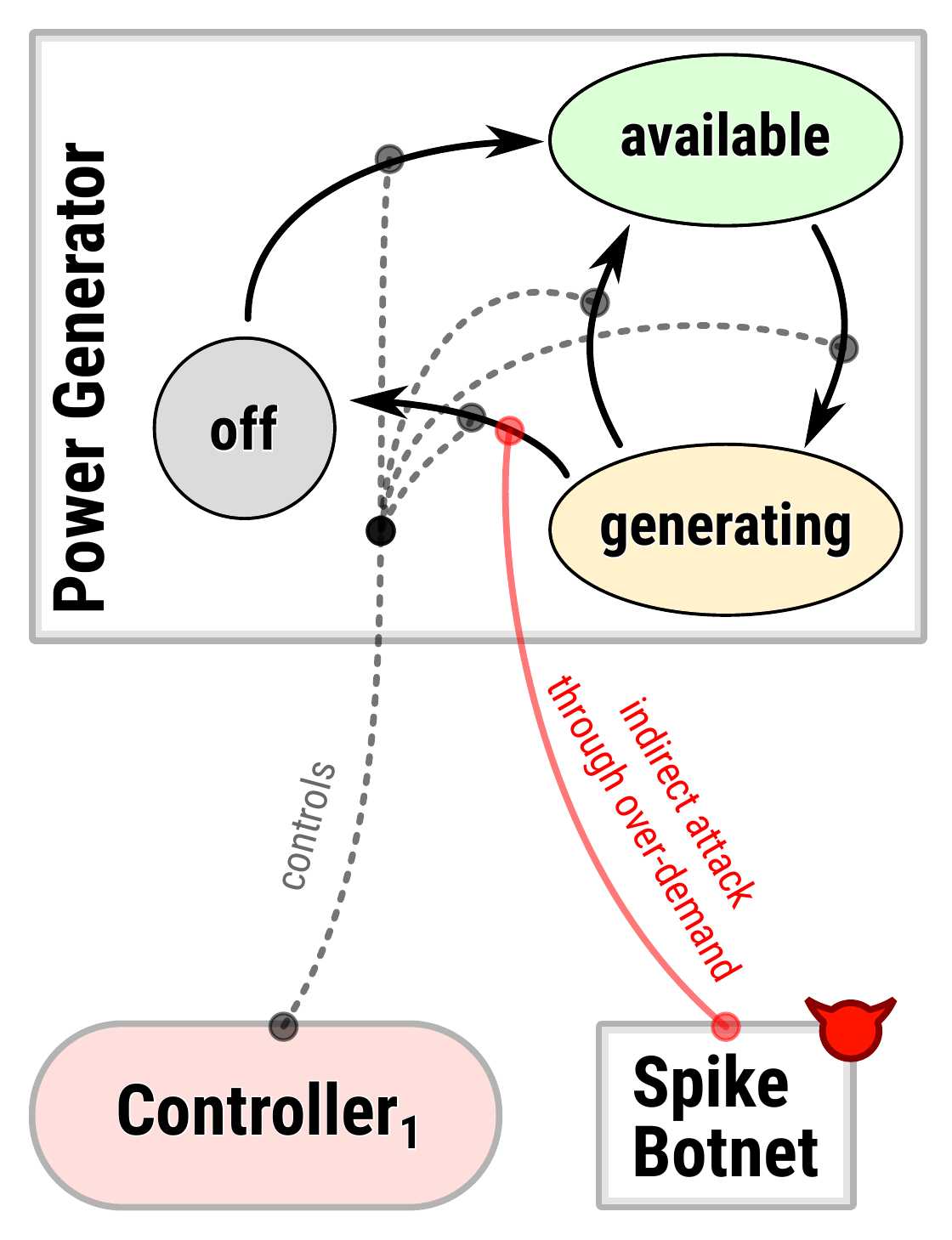}
  \caption{Attack-defence game between controller and attacker.}
  \label{fig:controller-attacker-powergenerator}
\end{subfigure}%
\begin{subfigure}{.6\textwidth}
  \centering
  \includegraphics[width=.9\linewidth]{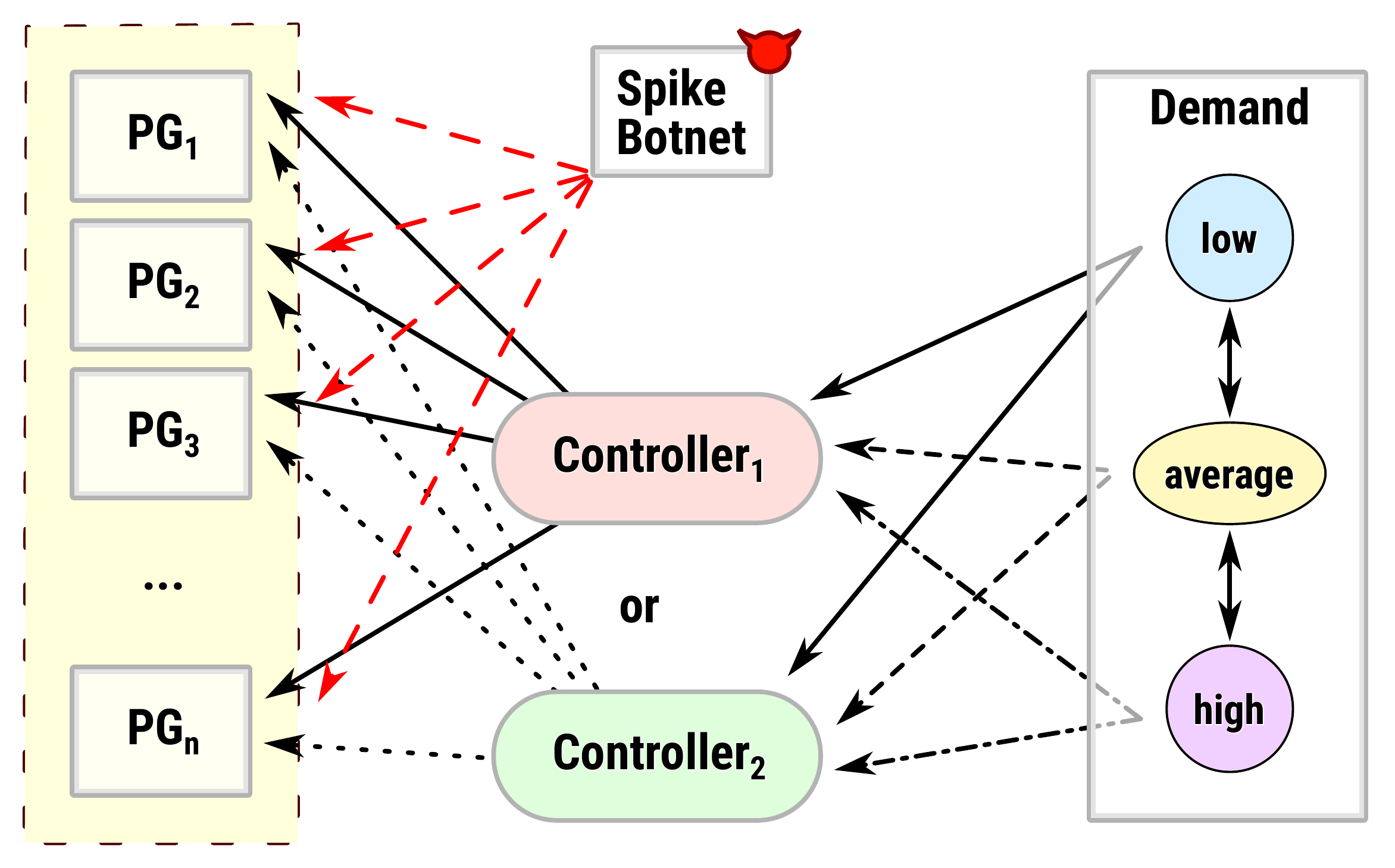}
  \caption{Different controller strategies for the same power grid.}
  \label{fig:controllers}
\end{subfigure}
\caption{The controller and the attacker role in our problem formalisation fine grained to the transitions of a PG (a) and coarse grained to the whole system with multiple PGs (b). We remark that the attack is not directly done to the power generator transition from {\em generating} to {\em off}, but the attack indirectly causes it through spike over-demand.}
\label{fig:controller-role}
\end{figure}
As the reader may notice, the controller of the power generator does not appear among the states.
This is not because there is a single PG, but by modelling choice that needs to refer to a more complex system with multiple PGs to be illustrated.

The basis for our model is presented in Fig.\ref{fig:controller-role}.
It is our intention to model a botnet (B) influencing the setting of power generators to let them go off, or detached.
In the normal operative state of the system, PG is available, then generating, or disconnected.
The demand operates at the medium value $m$, for some time going to low demand or high demand.
A controller dictates the response of the grid in case there is an excess of demand. 
It is a deterministic transition attempting to create an equilibrium in the grid.
In a system without an attacker, two different controllers may behave in the same exact manner, however under an attack the control strategy determines how effectively the load is re-balanced.
We define the optimum controller strategy as that one which minimises the time where a system is in a state of over supply or over demand.
What dictates the effectiveness of the controller is the responsiveness of the PGs, i.e the controller decides that a PG needs to be turned on to meet the demand, if the PG is a very slow one, this will lead to large amounts of time offline.
By modelling different controller strategies, one can easily envision that an optimal controller can be selected against a specific attack and under a specific load.
The less trivial research question is whether an optimal strategy can be found to optimise the power supply for a specific grid or CPS.
We investigate this problem by modelling the power supply in Sec.~\ref{sec:model}, and look at the way three different controller strategies impact the effectiveness of a spike Botnet attack.
We could potentially have a set of PG to work with (Nuclear, Gas, Electricity, Wind, and so on) where the Spike Botnet engages in an attack-defence game to roughly estimate the state of the controller to direct decisions on when to switch devices on or off.

\subsection{Energy Supply Demand Trade-off Model}

Our work focuses on modelling the balance of energy supply and demand. 
To do so we have created three entities:
i) {\em suppliers}, power generators mimicking different types of plants typically attached to the grid,
ii) {\em consumers}, modelled as average values of energy demand per hour, they also are subject to positive and negative variations across time, and
iii) {\em spike botnets}, a large quantity of compromised IoT devices, they might be thought as high wattage devices such as water heaters or air conditioners that can be synchronously turned on or off, they produce sudden spikes or drops of energy that unbalance the grid and trigger the automatic security disconnection mechanism of the PGs from the grid.
One way to mitigate disturbances is to employ \emph{load shedding} or \emph{tie-line tripping}, techniques employed to disregard incoming requests in order to maintain integrity and avoid breakages~\cite{dabrowski2018botnet}.

The way the demand supply trade-off is modelled is as following: the \emph{consumers} will have a certain energy requirement and the \emph{suppliers} will need to be turned on to meet the demand. However if the Botnet is successful in taking down a PG, there will be a temporary situation of under supply. 
To meet the demand, more PGs need to supply energy.
Their {\em responsiveness} is subject to several limitations.
First, a PG requires some time to be fully functional, especially if it gets suddenly detached from the grid; some are quite fast, while others are expected to be much slower, like nuclear power plants.
Second, the amount of PGs is finite; if the botnet takes down enough PGs whilst they are unable to reattach themselves to the grid, there will be no way for the suppliers to meet the demand.
Third, independently of how many PGs are taken down, at peak demand, the power grid might have very few spare resources, so if one PG is taken down at 18:00 during dinner time the impact might be a blackout for the whole grid.

\begin{figure}[!htbp]
	\centering
	\includegraphics[width=0.8\linewidth]{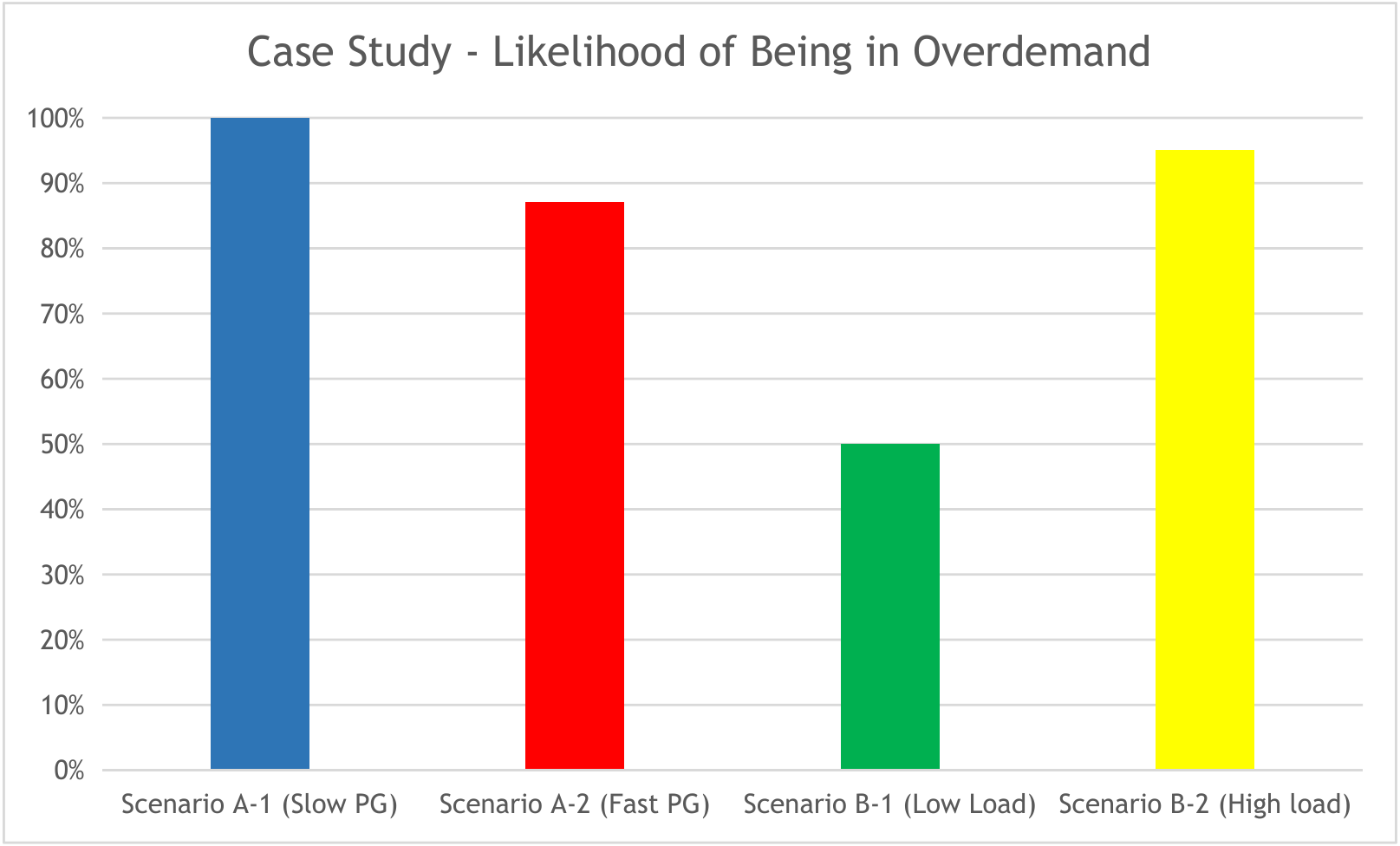}
	\caption{Powergrid reaction to botnet spikes in our case study scenarios}
	\label{fig:scenario}
\end{figure}

We show this balance in a toy example with two scenarios, Scenario A and Scenario B in Fig.~\ref{fig:scenario}, whose behaviour is as expected.
In this case study we show the impact of two different factors, responsiveness to the demand and magnitude of the demand, each placed in a low demand scenario and a high demand scenario.
Both scenarios are initially capable of meeting the demand, as there are plenty of available PGs; however, they react differently to the load caused by botnets' attack.

\begin{table}[htbpt]
\caption{How the PG's interact with the attacker in Scenario A-(1/2). The States are: \textbf{A} - \textit{Available}, \textbf{S} - \textit{Serving}, and \textbf{D} - \textit{Disconnected}. \textit{Sup}, is the supply of a single PG out of four and \textit{Att} is whether the Spike Botnet is on or not. The \textit{Demand} is fixed at 120 units. The values represent the rate at which a state transitions from a state to another (if 0 the transition doesn't exist).}
\begin{longtable}{|c|ccc|cc| c |c|ccc|cc|}
\cline{1-6}
\cline{8-13}
\multicolumn{1}{|c}{\multirow{2}{*}{\textbf{PG Slow}}} & \multicolumn{3}{|c|}{Transitions}   & \multicolumn{1}{c}{\multirow{2}{*}{Sup}} & \multicolumn{1}{c|}{\multirow{2}{*}{Att}}  &\ \ \ \  & \multicolumn{1}{c}{\multirow{2}{*}{\textbf{PG Fast}}} & \multicolumn{3}{|c|}{Transitions} & \multicolumn{1}{c}{\multirow{2}{*}{Sup}} & \multicolumn{1}{c|}{\multirow{2}{*}{Att}}\\\cline{2-4} \cline{9-11}

\multicolumn{1}{|c}{}  & \multicolumn{1}{|c}{A} & \multicolumn{1}{c}{S} & \multicolumn{1}{c|}{D} &\multicolumn{1}{c}{} & \multicolumn{1}{c|}{} & & \multicolumn{1}{c}{} & \multicolumn{1}{|c}{A} & \multicolumn{1}{c}{S} & \multicolumn{1}{c|}{D} &\multicolumn{1}{c}{}  & \multicolumn{1}{c|}{}  \\ \cline{1-6} \cline{8-13}
A & --  & 0.50 & 0  & 0 & Off & & A & --  & 1200 & 0  & 0 & Off \\
S & 0.50 & -- & 0  & 50  & Off && S & 1200 & -- & 0  & 50  & Off \\
D  & 0.25  & 0 & -- & 0  & Off && D  & 600  & 0 & -- & 0  & Off \\
A & -- & 0.50 & 1200 & 0 & On&& A & -- & 1200 & 120000 & 0 & On\\
S & 0.50 & --& 1200 & 50  & On&& S & 1200 & --& 120000 & 50  & On\\
D & 0.25 & 0 & --& 0  & On  &&D & 600 & 0 & --& 0  & On  \\
\cline{1-6} \cline{8-13}
\end{longtable}
\end{table}

In Scenario A, we model two systems A-1 and A-2 under a high load.
A-1 has very responsive PGs (such as gas plants), and A-2 has less responsive PGs. 
We showcase that in the system with faster response rate the time in which the system is in over demand is much lower.
On the other hand, due to slow startup times, in the second system the time in over demand will be much higher.

In Scenario B, we model two more systems, B-1 and B-2, each of the systems shows the same number of PGs, but while the demand of B-1 is high, the demand of B-2 is low.
We show that when the system is in a high demand period, an attacker is much more likely to disrupt the powergrid than if under a low demand period.
The scenario matches the values for Scenario A-1, but the demand is altered between B-1 and B-2 from 50 to 100. 
This case highlights the different situations that can take place if a Spike botnet were to target the power grid, in our in depth experiments mimicking the power usage of real world scenario we take this a step further and examine it on various different types of PG such as gas and solar plants.

\subsection{Power--Energy considerations}\label{subsec:energy-cons}

In energy terminology, \emph{Unit Commitment} (UC)~\cite{padhy2004unit} tries to find the least-cost dispatch of available generation resources to meet electrical loads.
In the past, different strategies were conceived to deal with UC related issues such as simulated annealing~\cite{simopoulos2006unit}, dynamic programming optimisation~\cite{papavasiliou2017application}, particle swarm~\cite{ting2006novel}, genetic algorithms~\cite{trivedi2016genetic,nemati2018optimization} or combination of those and other techniques.
UC is a relevant problem in energy as the demand/supply requirements are usually uncertain.

We are interested in working here with UC in an abstract way, representing supply and demand for CPS or smart grids.
There are several generating resources available for use by energy managers such as nuclear, thermal (using fossil fuels such as coal, natural gas or oil), or other biomass.
When deciding which power plant to turn on, several decision variables come into play such as generation level (in Megawatts, MW) and number of generating units that must be turned on.
A power plant employs different technologies to generate energy, for instance, nuclear based have to be turned on and then cooled down, tasks that usually take considerable time.
Other sources such as solar panels on the other hand have different maintenance peculiarities than those of, for example, wind turbines.
Lastly, when broken, each system would involve different resources and equipment, which translates to greater time not producing energy which impacts the grid in its entirety.

\begin{figure}[!htbp]
	\centering
	\includegraphics[width=0.7\linewidth]{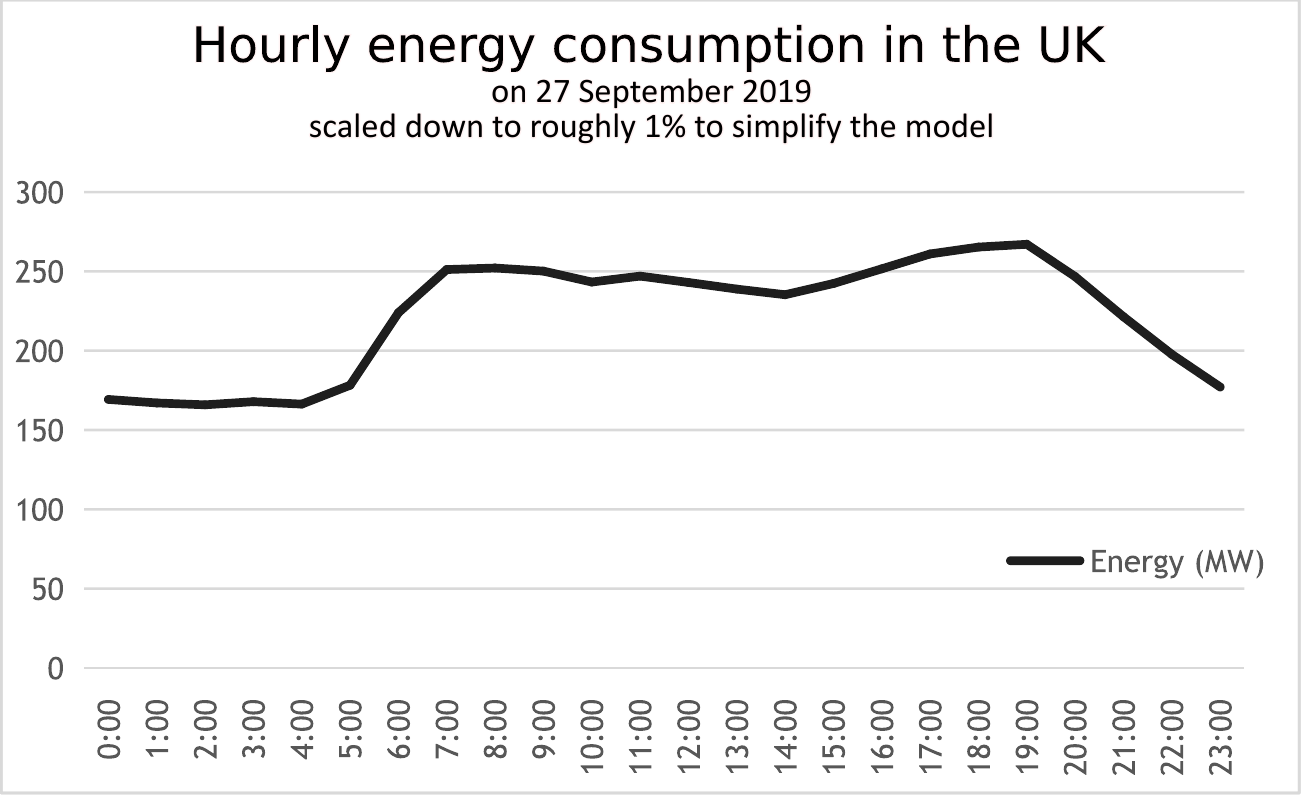}
	\caption{Usage data for the UK, scaled down to about 1\%, in MW, across 24 hours on Friday 27 September 2019 -- data obtained from \textit{Balancing Mechanism Reporting Service} (BMRS)~\cite{bmrs} }
	\label{fig:dataUsage} 
\end{figure}

Another source of concern is directed towards hourly \emph{fluctuations} (used interchangeably here as either \emph{surges} or \emph{peaks}) that may be present in a day.
These differences in demand-supply, if greater or lower than specific thresholds (defined by energy operators) may cause severe damage to the grid and or its turbines (if used), even sometimes causing permanent damages which impact projected energy yield.
Our model captures the so called \emph{sudden} surges to the energy grid, where infected high-wattage appliances synchronise their operations to trigger disruptions.
Fig.~\ref{fig:dataUsage} shows the daily energy usage for the United Kingdom scaled down to roughly its 1\% (e.g. Glasgow's population) with noticeable peaks in demand according to specific times of a given day\footnote{Original data is available online from the balancing mechanism reporting service in the UK~\cite{bmrs}.}.
We remark that we scaled down the data to allow for smaller state space in the model.

Fig.~\ref{fig:dataUsage} highlights an expected usage trend, with more power demand at times such as breakfast and dinner and a sharp decrease when (the most) people are asleep.
These are known trends and therefore power supplies are build to cater to them.
It is noticeable that the load is distributed in a way that if there is a reasonably expected raise in consumption at any point along the line the load controller will be able to handle it.
This \emph{tolerance} value is what the attacker has to outmatch in order to cause the disruption, perhaps unexpectedly, this means unexpected above normal usage is much more damaging than just mass usage. 
Through our model we highlight this by showing that the most impactful attack times (varying with the different power plants), may not be the ones of most usage.

In our setting we are dealing with a load-balancing system that tries to maintain the equilibrium to keep the frequency to 50 MHz; this is done balancing the demand (produced by users turning on their devices whichever they are) and the supply (the set of power generators required to meet the demand).
We are combining this balance with a control strategy, where we mimic the manager's decisions as to which power generator should be prompted at specific times to cope with demand.
These decisions address the fact that under surges or spikes it could be possible to increase the security level if the right measure is taken at the right time, working as a protection against those sorts of disruptions.
There are many ways to find reasonable forms of mixing which power to be switched on or off at any moment, and a common strategy is to use greedy algorithms for selection.
At the core of these algorithms it resides the fact that a prompt reaction should be deployed as soon as possible, so the next available power plant should be turned on adding power to the infrastructure so it can handle the demand accordingly.

\section{Cyber security model applied to CPS}\label{subsec:cps-model}

\begin{figure}[htbp]
	\centering
	\includegraphics[width=0.95\textwidth]{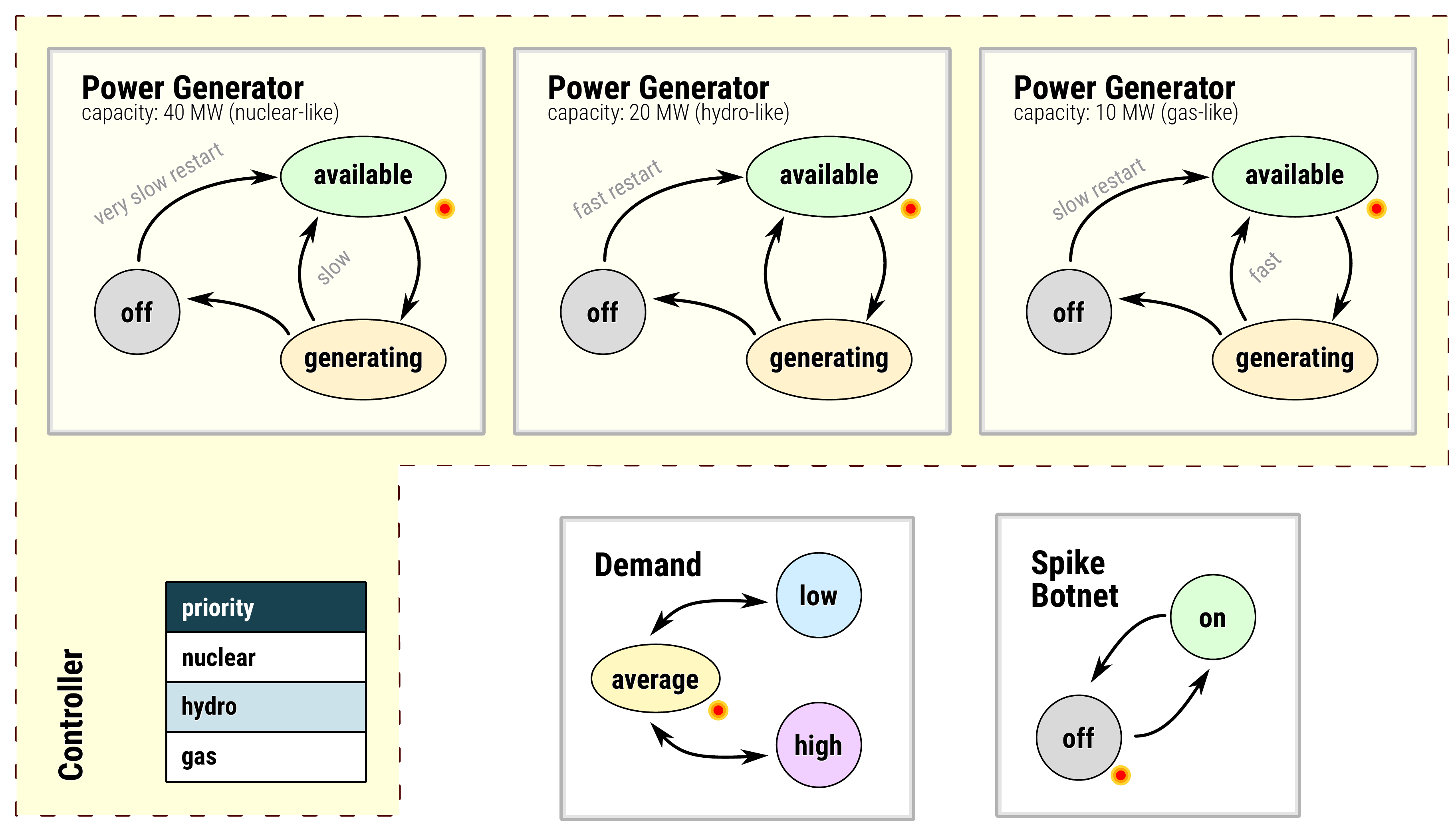}
	\caption{Model and PRISM modules, representing Nuclear, Hydro and Gas power; the demand, along with and variations from the expected value; and our designed attacker controlling a percentage of the systems devices.}
	\label{fig:model}
\end{figure}

In previous sections we discussed the problem from a general perspective.
For our experiment we focused on a scenario involving somehow realistic implementations of power generators and we adopt the real demand on a specific day in the UK.
For the model discussed here, we are interested in using CTMCs due to the required dynamics for our cyber-security problem.
Our modules are illustrated in Fig.~\ref{fig:model}: they consist of {power generators} PG (nuclear N, hydro H, and gas G), the {demand} D, modelling the supported threshold to try to withstand disruption case demand $\approx$ supply, a {controller} C, greedily selecting which power generator to use prioritising next according to design decisions, and finally the set of {IoT devices} I (the {Spike Botnet} in the figure), symbolising infected high-wattage components under the control of adversaries.
These choices were made to highlight a variety of setups mimicking a modern smart grid.

We are considering that each PG has its distinctive design possibilities, usually dealing with high loads of energy in different ways.
For our specific case study we are interested in how the different prioritisation of PGs could potentially affect the availability to cope with attacks.
In Fig.~\ref{fig:model},
we adopt a tailored notation to represent our modelling choices, for instance, initial states are marked with a \emph{small dot} whereas rates are shown as textual descriptions decorating some transitions (e.g. \emph{slow, fast} and so on).

All PGs start in \emph{available} state when ready to supply, in the \emph{generating} state when supplying to the grid, or in \emph{off} state when offline.
We have added a module Controller to cope with the supported variation, parameterised to withstand increments or decrements of $\pm1\% W$.
Finally, the infected devices (botnets) are either \emph{on} or \emph{off}, modelling either an attack in place or inactive, respectively.
In our model, the Botnets are composed by unique devices that are only controlled by the Botnet and do not represent daily usage.
The state delays (they will be converted to rates due to our CTMC representation) for the model are shown in Table~\ref{tab:params}.

\begin{table}[htbp]
\centering
\renewcommand\thetable{2}
\caption{Parameters for the model in PRISM and total of instances for every PG.}
\label{tab:general_params}
\begin{longtable}{|l|c|c|c|} 
 \hline
 \textbf{PRISM Module} & \multicolumn{2}{|c|}{\textbf{Observation}} & \textbf{\# of instances} \\ \hline 
 Nuclear (N) & \multirow{3}{*}{Generates} & $40$MW each & $4$ \\ 
 Hydro (H) & & $20$MW each & $5$ \\ 
 Gas (G) & & $10$MW each & $6$ \\ \hline
 Controller (C) & \multicolumn{2}{|c|}{$\pm$1\% tolerated deviation from normal} & -- \\ 
 IoT Devices (I) & \multicolumn{2}{|c|}{up to $30$\% expected wattage} & -- \\ \hline
 \multicolumn{3}{|r|}{\textbf{Total:}} & $320$MW\\ \hline
\end{longtable}
\end{table}

We changed the PGs characteristics through fixed parameters that map their behaviour.
We assume that the power generators' capacity does not depend on the hour of the day, even if it may not be the case, i.e. in night hours we obviously expect solar generator to have reduced supply capacity.
This can anyways be modelled, as we run experiments by per hour, and this time can be reduced to be more fine grained.
The Table~\ref{tab:general_params} shows the parameters we used, where the number of module instances mimic the specific scenario of matching the power supply for a location populated roughly as much as Glasgow is.
We do not realistically refer to Glasgow's power plan scenarios, only to the expected demand of roughly its population. 
The supply will always cater to the daily demand, however the IoT devices may still cause spikes to offset the load management and break PGs.
It is reasonable to consider that attackers may select regions where the difference between supply and demand are close.
In low consumption regions (in Fig.~\ref{fig:dataUsage} they correspond to early morning) or high (start of the day at 8:00 and then at 20:00), the power plants are taking decisions as whether to increase or reduce the power supply to meet the demand.
Our modelling approach focus on closely inspecting those time spans, aiming to evaluate which power plant configuration would be best suited to be adopted in case the infrastructure was to be targeted for attacks. 

\noindent
We are considering the following scenarios for our analysis:
\begin{enumerate}
    \item \textbf{NO-ATTACK:} normal grid operation. It is expected to be in \emph{over supply} state most of the time;
    \item \textbf{ATTACK:} in this scenario we are modelling the coordinated effort to disrupt the grid. 
    The attackers could profit from the peak hour knowledge to direct their efforts. 
    We are varying Controller to prioritise which PG would be triggered first to investigate the possible mitigation strategies as follows:
    \begin{enumerate}
        \item \textbf{ATTACK-N:} Controller prioritises first on Nuclear, then Hydro, then Gas.
        \item \textbf{ATTACK-H:} Priority on Hydro, then Nuclear, finally Gas.
        \item \textbf{ATTACK-G:} For this one, first turn on Gas, then Nuclear, finally Hydro.
    \end{enumerate}
\end{enumerate}

For our four experiments we have instantiated a total of $17$ modules, addressing in terms of PG a total of $320$ MW.
This power corresponds to a grid that could be deployed in a region to serve the power needs of the whole of Glasgow (population $\approx598,830$ people).
The reachable state space of our NO-ATTACK model is around $\approx300.000$  whereas in the ATTACK-* models, it sums to $57$ million (with the modified Controller policy and attack botnets).
We computed the model properties related to the probabilities of the system being \emph{over supply}, \emph{in equilibrium} and \emph{over demand}. In terms of rates, we are considering the values of Table~\ref{tab:params}.
\begin{table}[htbp]
\centering
\renewcommand\thetable{3}
\caption{Parameters of individual modules used in the model (time scales in minutes (m) or seconds (s).}
\label{tab:params}
\begin{longtable}{|c|c|c|c|}
\hline
\multirow{2}{*}{\textbf{Module}} & \multicolumn{2}{c|}{\textbf{State}} & \multirow{2}{*}{\textbf{Time}} \\ \cline{2-3}
& \textbf{From}     & \textbf{To}     &   \\ \hline\hline
\multicolumn{1}{|c|}{\multirow{3}{*}{\textit{Nuclear}}} & Available & Serving &  30s   \\ \cline{2-4} 
\multicolumn{1}{|c|}{}   & Serving & Available & 40m  \\ \cline{2-4} 
\multicolumn{1}{|c|}{}   & Serving & Offline & \textless 1s  \\
\hline
\multirow{3}{*}{\textit{Gas}}  & Available & Serving & \textless 1s   \\ \cline{2-4} 
    & Serving & Available &  30s   \\ \cline{2-4} 
    & Serving & Offline & \textless 1s   \\ \hline
\multirow{4}{*}{\textit{Hydro}}  & Available & Serving &  \textless 1s \\ \cline{2-4} 
    & Serving & Available & \textless 1s  \\ \cline{2-4}
    & Serving & Offline & \textless 1s  \\ \cline{2-4} 
    & Offline & Available &  20m  \\ \hline\hline
\multirow{4}{*}{\textit{Demand}}  & Normal & Negative & 5m  \\ \cline{2-4} 
& Negative & Normal & 1m  \\ \cline{2-4} 
 & Normal & Positive & 5m  \\ \cline{2-4} 
& Positive & Normal & 1m  \\ \hline\hline
\textit{IoT Devices }  & Off & On & 1m   \\ \cline{2-4} 
    \textit{Spike Botnet} & On & Off & 1m   \\ \hline
\end{longtable}
\end{table}
\section{Results}\label{sec:results}

Among the several interesting analysis that could potentially be performed using our model, we direct the focus to the probability for the over demand when a spike is successful.
In such a case, we consider an attack successful if the spike makes a PG go offline, after having caused an excess in demand. 
The controller prioritises the order of activation of PGs (we assume all are under its control).
If all preferred controllers are on, an alternative type will be activated. 

\begin{figure}[htbp]
	\centering
	\includegraphics[width=0.8\textwidth]{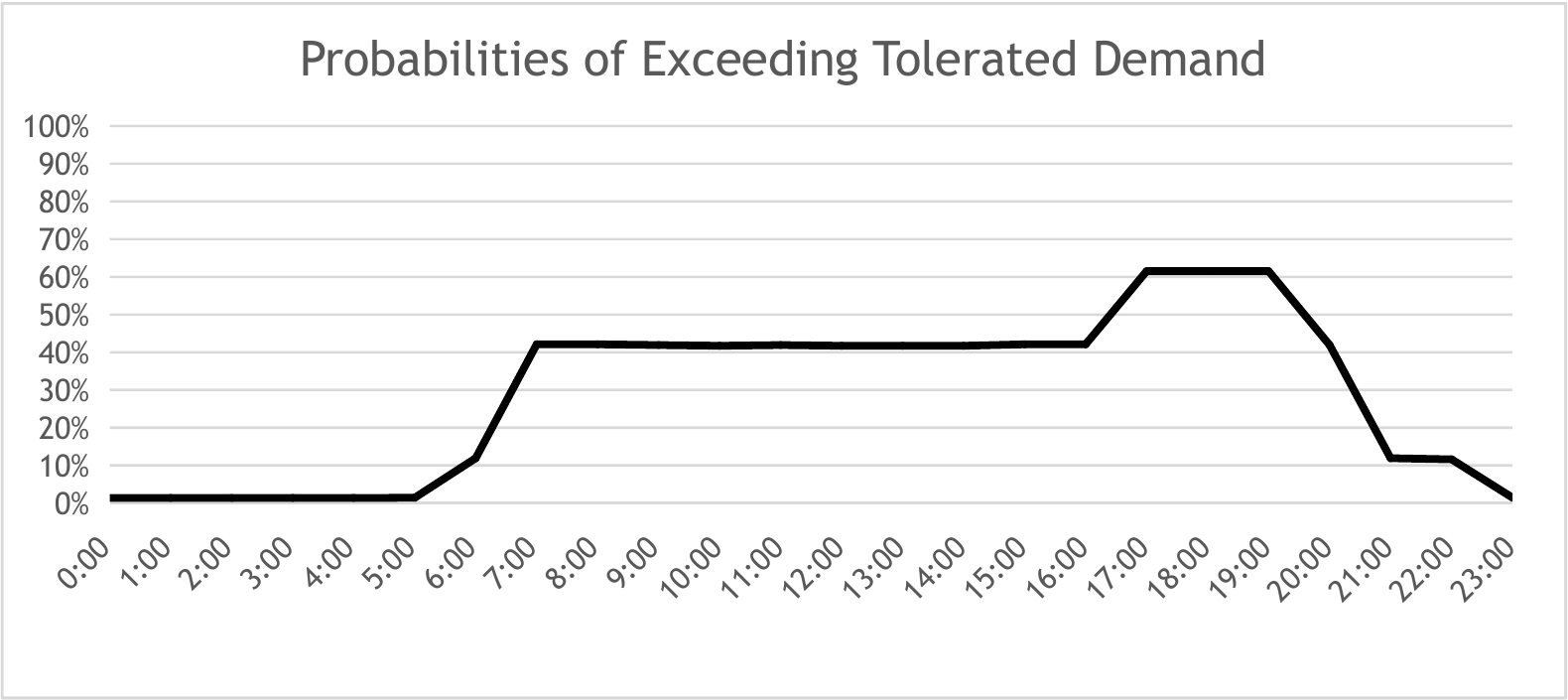}
	\caption{Probabilities for demand raising above tolerated values, in a day to day scenario with our PG setup.}
	\label{fig:probabilityUsage}
\end{figure}

Our results show key times in which the demand may exceed the current supply, attackers may use those ranges as clear opportunities for disruptions.
We have four main scenarios (NO-ATTACK, ATTACK-N, ATTACK-H and ATTACK-G) as described in the previous section, each one comprising one hour of a day, totalling $4\times24=96$ PRISM models\footnote{We have used PRISM version $4.5$.}.
The attack scenarios resulted in a state space of an average of $57,264,556$ million states, so the experiments were run in parallel on a multi-core server with 16 processing units.
The model parameters were tuned to represent the daily energy usage of an area populated roughly as much as 1\% of the UK (like a big city such as Glasgow in Scotland), with mean usage and expected variations at each hour of the day.
In our first experiment, we produced the baseline usage to find the likelihoods of exceeding the demand (compare with results in Fig.~\ref{fig:probabilityUsage}), this allowed us to evaluate the expected behaviour of the system without a spike botnet.

\begin{figure}[htbp]
	\centering
	\includegraphics[width=0.8\linewidth]{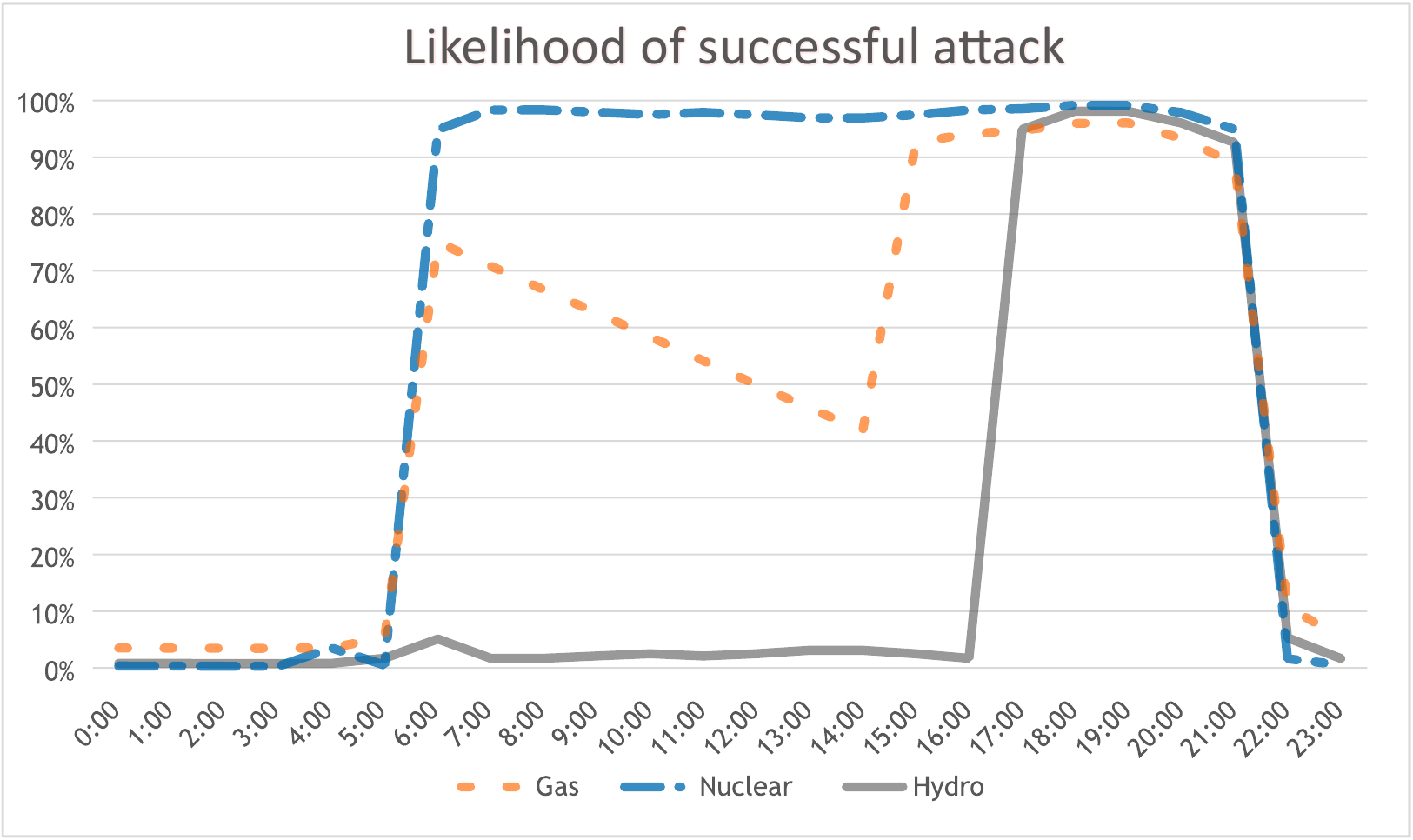}
	\caption{Likelihood of blackouts caused by spike botnets and different control strategies.}
	\label{fig:attackResults}
\end{figure}

The next three experiments each modelled a different power prioritisation to meet the demand, this allowed us to investigate which power generator type performs best under the threat of a spike botnet, interestingly there are quite a few differences as highlighted by Fig.~\ref{fig:attackResults}.

Fig.~\ref{fig:probabilityUsage} shows our demand probabilities for a day.
It is noticeable the close relation of the demand with the daily power consumption of Fig.~\ref{fig:dataUsage}.
The difference is that, with this graph, stakeholders may also inspect the probability that the demand will exceed and then anticipate load-changing opportunities of attacks.
The lowest probabilities are during early morning (before 6:00) and after 10pm.
During the day, there is a considerable probability (around 40\%) of finding the system under high load whereas the chances increase to approximately 60\% by 5:00 to 19:30, slowing down from then until 23:00.

Fig.~\ref{fig:attackResults} shows the results from our attack models.
As expected, peak hours are more susceptible for spike attacks; however, it is possible to infer the probability of the system where it is in over demand, information that may be used to have auxiliary power generators readily available to avoid energy interruption in the event of attacks.
The different controllers actually performed significantly differently with respect to security.
Prioritising Hydro led to the scenario where an attack is the least likely to succeed across the day.
Nuclear, on the other hand, performs rather poorly, this is due to it's slow transitions from \emph{serving} to \emph{available} and due to it's inability to recover after a spike attack (within 1 hour).
Finally, prioritising gas shows more areas of attacks than hydro.
It also reaches a slightly lower height at peak time of 18:00 (peak time of success for all attack scenarios), rendering it better over hydro power for that particular time period.

We stress that the numbers we used to model the PGs are partially reflecting the reality, as we borrowed them from what specifications we could find, but they may not be as realistic as wanted by experts in real specific domains.
Building on our example, one could always re-run the experiments with different characteristics and settings, to better reflect their scenarios of interest.
It would then be possible to compute the probability of being under high or low demand given these new parameters and devise reasonable countermeasures to avoid peak attacks.
These peaks are not necessarily cause by malicious attackers, but may happen due to other benign causes.
In Brazil in 2007, for example, hackers were promptly blamed as the cause of a major blackout, but further investigation concluded that some faulty equipment triggered a cascading failure from a poorly maintained systems\footnote{Official document: \url{http://www.aneel.gov.br/cedoc/adsp2009278_1.pdf}}.

\section{Conclusion \& Future Work}\label{sec:final}

The pervasive nature of CPS in real world applications mandates system designers to address security concerns more deeply.
Security related problems directly impact critical infrastructures, requiring the installation of additional software patches, inevitably causing monetary losses to stakeholders.
Power reliant stations are concerned with the responsible provision of energy to customers, since disruptions may impact costs and more serious outcomes (human lives).
We stress the fact that properties such as integrity, availability, reliability, dependability~\cite{avizienis2004basic}, plus performance, safety and security are crucial quality properties to deliver adequate service levels and enhance trust to users.

The problem addressed in this work tackled {Unit Commitment} and presented models to enhance power plant responses to active attacks targeting the energy infrastructure.
We have built a CTMC mapping to a power grid mixed with a network of CPS for a set of scenarios.
We have considered situations for normal operation (assuming no attacks) in conjunction with scenarios where attackers acting in coordination could take advantage of infected CPS to cause electric peaks or surges, which in turn would increase or decrease power consumption in relatively short time periods.
We have also presented scenarios where power managers could enact mitigation policies, e.g. choosing other power sources or disconnecting parts of the grid to meet demand.
The model described here could also be applied in another mechanism where managers could differentiate localised abnormal usage from directed attacks and take measures to avoid grid disruptions more effectively.

Our results show the numerical impact of addressing mitigation strategies to avoid peaks using CTMCs to capture intricate power related behaviours.
We show how different power generators produce different results under attack and also that a intelligent controller could be designed to alter the priority of power generation and somewhat mitigate these kinds of attacks.

As next work, we are envisioning to extend the model to address more attackers with different patterns of behaviours as well as adding costs and rewards to states deemed important for later analysis.
We could also add extra states to model the activation of other external power sources such as secondary or even tertiary mechanisms and their relation to attacks.
The model could also be scaled in terms of different CPS types and wattage needs, where we could evaluate the impact on new security measures to avoid powering up new plants or blacking out major grid portions in response.

It is also our intention to work with simulation models mapping and addressing some of the concerns provided here, where we could enrich it with more fine tuned behaviours to capture other characteristics of power plants as well as other adversary profiles.
It is also worth mentioning that in the work concerning {GridShock}~\cite{dabrowski2017grid}, the authors suggest other power plants start up mechanisms in the event of sudden disconnections, e.g. \emph{black restart}, where portions of the grid are turned on each time.
We could analyse these procedures in combination with the notion that cyber attackers may profit from such slow restarts to disrupt the grid and stop energy to be distributed over the grid.

Finally, our model could be instantiated with other power sources (e.g. wind mills or attaching electric vehicles), where we could investigate other relationships between power plant configuration mixing and cyber-security attacks altogether.
In retrospect, our approach has the potential to suggest means to improve their operations as well as reasonable design approaches towards more reliable uses of energy sources as well as mitigation strategies to improve overall operational requirements.

\section*{Acknowledgements}
This research is supported by The Alan Turing Institute and an Innovate UK grant to Newcastle University through the e4future project, Arm Ltd. and EPSRC under grant EP/N509528/1, as well as the Active Building Center under grant EP/S016627/1. %

\end{document}